\documentclass[useAMS,usenatbib]{mn2e}
\usepackage{graphicx}
\usepackage{amsmath}
 \usepackage{color}
 \usepackage[normalem]{ulem}

 \title[Testing the rotating  lighthouse model  with the double pulsar]{Testing the rotating  lighthouse model  with
the double pulsar system PSR J0737-3039A/B}
\author[Z.-X. Liang, Y. Liang, \& J. M. Weisberg]{Zhu-Xing Liang$^{}
$\thanks{zx.liang55@gmail.com (LZX); jluliangyi@gmail.com (LY)} and Yi Liang$^{}$\\
$^{}$18-4-102 Shuixiehuadu, Zhufengdajie, Shijiazhuang, Hebei 050035,China\\
\newauthor
Joel M. Weisberg$^{}$\\
$^{}$Department of Physics and Astronomy, Carleton College, Northfield, Minnesota
55057, USA}

\begin{document}

\date{}

\pagerange{\pageref{firstpage}--\pageref{lastpage}} \pubyear{2002}

\maketitle

\label{firstpage}

\begin{abstract}

Each of the two pulsars in the double pulsar PSR J0737-3039A/B
system exhibits not only the pulses
emanating from itself, but  also displays modulations
near the pulse period of the other.
Freire et al. (2009, MNRAS, 396, 1764) have put forward a technique
using the modulation of B by A to determine the
sense of rotation of pulsar A relative to its orbital
motion, among other quantities.  In this paper,
we present another technique with the same purpose. While the Freire et al.
approach analyzes pulse arrival times, ours instead
uses periods or frequencies (their inverses), which can be experimentally determined via power
spectral analysis similar to that used in pulsar searches.  Our
technique is based on the apparent change in spin period of a body when
it is measured from an orbiting platform (the other pulsar), and is shown to be entirely
analogous to the difference between the sidereal and solar spin period
 of the Earth (i.e., the  sidereal and solar day).  Two benefits of this approach
 are its conceptual and computational simplicity.  The direct detection of spin
 with this technique will observationally validate the rotating lighthouse model
 of pulsar emission, while the detection of the relative
 directions of spin and orbital angular momenta has important evolutionary
 implications. Our technique can be used on other binary systems exhibiting
 mutually induced phenomena.

\end{abstract}


\section{Introduction}

The double pulsar system PSR J0737-3039A/B was discovered by
 \citet{Burgay03} and \citet{Lyne04}. This system consists of a 22-ms
pulsar (hereafter A) and a 2.8-s pulsar (hereafter B) with an
orbital period of 2.4 hours. This discovery has provided a laboratory
for the study of relativistic gravity and gravitational radiation
 \citep{KramerWex2009}. The system has several strange features that
 challenge the current
understanding of pulsars and provide an uncommon opportunity to
improve pulsar theories. One of the most interesting properties is the
observed modulation
of each pulsar's signal
by the energy flux from the other, as evidenced
by each pulsar's modulation period being approximately equal
to the other pulsar's pulse period \citep{McLaughlin04a,McLaughlin04b}.
\citet{Freire09a}
proposed  a technique for analysing the arrival times of the pulsars'
pulses and their mutual modulations which could yield the sense of rotation of
each pulsar with respect to its orbital motion, among other quantities.
In this paper, a complementary technique is presented
with the same objective, but using measured periods rather than arrival
times.   The principal benefit of our approach is that it is simpler and more intuitive.  If the validity of either of these techniques is
confirmed, not only will new insights be gained but also the
correctness of the lighthouse model will be empirically assessed
 beyond dispute.
\begin{table*}
 \centering
 \caption{Earth-Sun  and double pulsar  analogies. All quantities are defined as magnitudes.}
 \begin{minipage}{184mm}
\begin{tabular}{lcc}
\hline
Rotating object:& Earth $(\earth)$&Pulsar A \\
\hline
Orbital period, Mean orbital frequency  &   $P_{\rm orb} (\equiv$ year), $f_{\rm orb}$ &  $P_{\rm orb}, f_{\rm orb}$\\
\hline
 \hspace{0.4in} Instantaneous  orbital  frequency   &   \hspace{0.7in} $f_{\rm orb}(t)$  &  \hspace{0.43in}  $ f_{\rm orb}(t)  $    \\
\hline
Rotating object's  sidereal spin period, frequency  &   $P_{\earth,0} (\equiv$ sidereal day), $f_{\earth,0}$ &   $P_{\rm A,0}, f_{\rm A,0} $  \\
\hline
 Rotating object's spin period, frequency  (measured at other body) &   $P_{\earth\ \rm at\  \sun} (\equiv$solar day), $f_{\earth\ \rm at\  \sun}$ &   $P_{\rm A\ at\  B}, f_{\rm A\ at\  B}$  \\

$\equiv\ $Modulation period, frequency at other body due to rotating object   & $\equiv P_{\rm m\ at\  \sun\ due\ to\ \earth},$  &   $\equiv P_{\rm m\ at\  B\ due\ to\ A}, $ \\
                         &\ \ \ $ f_{\rm m\ at\  \sun\ due\ to\ \earth} $&  \ \ \ $f_{\rm m\ at\  B\ due\ to\ A}$  \\

\hline
\end{tabular}
 \end{minipage}
\label{tab:one}
\end{table*}

\section{Determining the presence and sense of
 rotation of A, and analogies to the Earth-Sun system}

Our method is identical in principle to the determination of the difference
between two measurements of the Earth's spin period -- its
sidereal and solar day.  The difference between
the two periods is caused by the kinematic effect of Earth's
rotation and revolution. Because the earth rotates and revolves in
the same (``prograde'') sense, the solar day is about 4 min longer than the
sidereal day. If the earth rotated in the opposite (``retrograde'') sense relative to
its orbit, the solar day would be about 4 min shorter than the
sidereal day. A similar relationship should be seen in the PSR
J0737-3039A/B binary system if the lighthouse model is correct, where
the apparent period of one pulsar's pulses measured at the other pulsar
-- the other pulsar's ``modulation period'' --  will be longer or shorter than
 the first pulsar's sidereal period if the first is a ``lighthouse''
rotating in the same or opposite direction as it is orbiting.

Although our technique can in principle
be used to determine whether or not both A and B are
 spinning and the sense of such rotation with respect to their orbital motion,
 we focus only on determination of the spin of A  in this and next sections, in order to avoid confusion.
 Then Earth's  sidereal day, $P_{\earth,0}$, is analogous to
A's sidereal rotation period, $P_{\rm A,0}$.     Similarly, Earth's solar day,
$P_{\earth\ \rm at\  \sun}$, is analogous to
the modulation period  in B's, signal, $P_{\rm m\ at\  B\ due\ to\ A }$.
Table 1 further illustrates
the analogies between the Earth-Sun  system and the  double pulsar system,
in terms of  the various periods and their corresponding frequencies $f=1/P$.
In what follows, we will generally use spin and orbital  frequencies instead of
periods, because the frequency calculations are simpler.

To further understand the
relationships among the rotation, revolution and modulation signals, consider
the following suppositional extreme case. Assume
that  the spin periods of A and B are both equal to the
orbital period and that both stars rotate in the same sense as their orbital motion.
Then their emission signals will not modulate
each other because each star can receive only steady energy
fluxes from the other star.
In any other case, it is clear that modulation of one pulsar's signal by the other
is possible, and that the period of modulation will be affected  not only by  the
rotation period of
the pulsar causing the modulation but also its orbital period and rotation sense.

From  the analysis above, we have shown that as long as we can
measure  the modulation frequency of B's signal, we can
determine the presence and sense of rotation of A.

\section{frequency determinations}

When a radiotelescope points to the system PSR J0737-3039A/B, up to four
periodicities may
 be received simultaneously; namely A's
 and  B's direct and modulated signals. They will be intermingled at the radiotelescope since
they are spatially indistinguishable at the solar system. In this section, we focus on the  signals originating at A.
(See Sec. \ref{sec:rotB}
 for a discussion of  signals originating at B.)

\subsection{Signal paths originating at A}
Signals  are observed from from pulsar A via  two different paths: (1) the ``direct beam''
 travelling directly from  A
to the solar system; and (2) the ``two-legged'' beam emitted by A that first impinges on B,
thereby modulating B's beam every time that A shines on B; with the modulated signal
then travelling from B to the solar system. Consequently, the modulated signal  encodes information   concerning \textit{both} B's and A's pulses.

\begin{figure}
\includegraphics[trim=18mm 0 0 0,width=80mm]{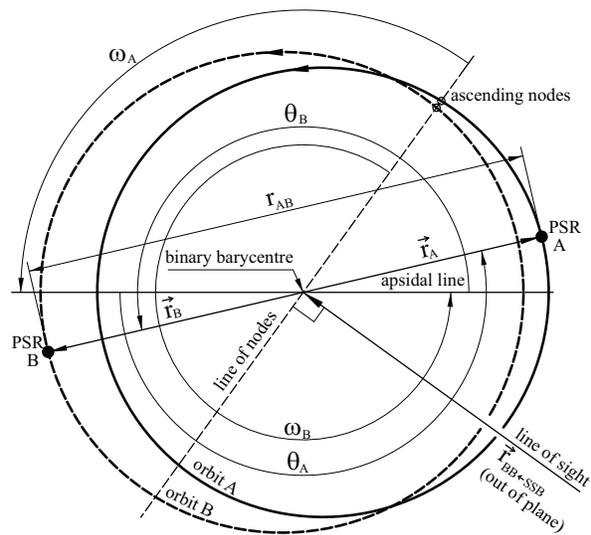}
\caption{Orbits of PSRs A and B  about the binary barycentre. The {\it{direct}} signal path discussed in the text is emitted from pulsar A antiparallel to the line of sight vector; while  the ``two-legged''
signal path also starts at A but then travels to B before travelling antiparallel to the line of sight vector. }
\label{fig:orbits}
\end{figure}

Details of the orbital geometry are shown in Fig. \ref{fig:orbits}.
Note that it is the first leg of the two-legged path which provides the capability to distinguish the direction of A's spin, if any; because it enables us to measure A's pulse period with respect to a another vantage point (namely pulsar B's), in addition to our own.

\subsection{Frequency calculations}

It will be adequate for our purposes to ignore  kinematic modifications
to source frequencies  beyond  order $v/c$, leaving only the first-order Doppler shift.  The
Doppler shift
 modifies   $f_{\rm X, 0}$, the emitted frequency of source `X', leading to a
 received frequency $f_{\rm X, rcv}$ according to the following prescription:

\begin{equation}
 f_{\rm X, rcv} =       \left[ 1+\frac{   \vec{  \mathbf v}_{\rm X} \cdot
 \hat{\mathbf r}_{\rm X \leftarrow rcv} } {c}\right] ^{-1} \times\ \  f_{\rm X, 0}\ ,
\label{eqn:dopplerexplicit}
\end{equation}
where the dot-product projects   the  velocity of source `X'  with respect to  receiver `rcv' ,
$\vec{\mathbf v}_{\rm X} $, onto the
receiver-to-source-`X'   line-of-sight  unit vector $\hat{\mathbf r}_{\rm X \leftarrow rcv}$,  thereby
yielding the ``radial'' (with respect to the receiver)  velocity component.
In what follows, source `X'  may be PSR A or PSR B, and  `rcv'  may be located at the
solar system
barycentre, ssb; or even at PSR B (see below).  In each such case, the subscripts will be
replaced by appropriate symbols to denote the particular choices.

We define the ``Doppler factor'' {\bf{D}}[X, rcv] solely for compactness of notation:
\begin{equation}
{\mathbf{D}} [ {\rm X,rcv}] \equiv
\left[ 1+\frac{   \vec{  \mathbf v}_{\rm X} \cdot \hat{\mathbf r}_{\rm X \leftarrow rcv}     }{c}\right] ^{-1},
\label{eqn:dopplershort}
\end{equation}
such that Eq. \ref{eqn:dopplerexplicit} becomes
\begin{equation}
f_{\rm X, rcv}= {\mathbf{D}} [ \rm X,rcv]  \times f_{\rm X,0}.
\end{equation}
(The two arguments of {\bf{D}} denote the source and the receiver, respectively.)

\begin{table*}
 \centering
 \begin{minipage}{184mm}
  \caption{Magnitudes of frequencies $f (=1/P)$ on two signal paths for  A's emission.}

\begin{tabular}{@{}lccc@{}}

\hline
\multicolumn{4}{c}{\bf{Signal path 1: Direct beam from A   to  solar system barycentre (ssb)}}\\
\hline
Intrinsic spin frequency of A:               &  &  $f_{\rm A,0}$   \\
Apparent spin frequency of A at ssb, \\
\hspace{2.8cm} $f_{\rm 1,ssb}=f_{\rm A,ssb}$:        &  &   ${\mathbf{D}}  [ {\rm A,ssb} ] \times f_{\rm A,0} $   \\
\hline
\multicolumn{4}{c}{\bf{Signal path 2: Two-legged beam from A to B to solar system barycentre}}\\
\hline
&\multicolumn{2}{c}{\underline{lighthouse model}} & \underline{other model}\\
A's rotation sense relative to orbit: & prograde & retrograde & none\\
\\
Intrinsic modulation frequency of B, \\
\hspace{2.5cm} $f_{\rm m\ at\  B\ due\ to\ A}$:
&    $ {\mathbf{D}} [{\rm A,B}] \times  (f_{\rm A,0}-f_{\rm orb}(t) )\  $\footnote{\.The instantaneous
orbital  frequency $f_{\rm orb}(t) $ varies around the elliptical orbit:
 $f_{\rm orb}(t)=\frac{1}{ 2 \pi}\frac{d \theta(t)}{ dt}.$  For a circular orbit,
 $f_{\rm orb}(t) \rightarrow  f_{\rm orb}= 1 / P_{\rm orb}$. }
&   ${\mathbf{D}} [{\rm A,B}] \times (f_{\rm A,0}+f_{\rm orb}(t)  )\ ^a$  &
${\mathbf{D}} [{\rm A,B}] \times f_{\rm A,0}$ \\

Apparent modulation frequency of  \\
\hspace{1.5cm}B at ssb, $f_{\rm 2, ssb}=f_{\rm m, ssb}$:
& $ {\mathbf{D}} [{\rm B, ssb}] \times {\mathbf{D}} [{\rm A,B}] \times (f_{\rm A,0}-f_{\rm orb}(t) )$
& $ {\mathbf{D}} [{\rm B, ssb}] \times{\mathbf{D}} [{\rm A,B}] \times  (f_{\rm A,0}+f_{\rm orb}(t) )$
& $ {\mathbf{D}} [{\rm B, ssb}] \times  {\mathbf{D}} [{\rm A,B}] \times   f_{\rm A,0}$  \\

\hline

\end{tabular}
\end{minipage}
\end{table*}

\subsubsection{Radial velocity calculations}
\label{sec:radialvel}

There are well-developed techniques for determining the radial velocity of an orbiting
body X (the dot-product
appearing in Eqs. \ref{eqn:dopplerexplicit} and \ref{eqn:dopplershort}) in terms of
its orbital elements.  Following
 \citet{Freire01,Freire09b}, the radial  (with respect to the receiver)   velocity
 of object X, at any point in its orbit, is given by
 \begin{equation}
  \vec{  \mathbf v}_{\rm X} \cdot \hat{\mathbf r}_{\rm X \leftarrow  rcv}  =
  \frac {2 \pi}{ P_{\rm orb}  }\ \frac  { a_{\rm X} \sin i} {\sqrt{1-e^2} }
  [\cos (\omega_{\rm X} + \theta(t)) + e \cos \omega_{X}],
  \end{equation}
  where $a_{\rm X}$ is the  semimajor axis of the orbit of `X',  inclination $i$ is the angle
  between  a vector normal to the  orbital plane and
   $\hat{\mathbf r}_{\rm X \leftarrow obs}, e$ is the orbital eccentricity,  $\omega_{X}$ is the
   argument of periastron of
  X's orbit, and true anomaly $\theta(t)$ is the polar angle of `X' at emission time $t$, measured
  around the principal focus of its orbit, starting at periastron. Finally,
  in order relate all of these quantities explicitly to time, one must
   determine the value of $\theta(t)$ at a given emission time via the iterative
 ``Kepler's Equation''  technique, as given in    celestial mechanics texts [
 e.g.,  \citet{Roy05} ].  Note also that $\omega_{X}$
 is also a (slowly  and essentially linearly-varying) function of $t$, due to a  general
 relativistic phenomenon  that is easily accounted for.   Finally,
  for the special case of the relative orbit of PSR A about PSR B, Eq. \ref{eqn:dopplershort}  becomes

\begin{equation}
  {\mathbf{D}} [ {\rm A,B}] =
\left[ 1+\frac{   \vec{  \mathbf v}_{\rm rel,A} \cdot \hat{\mathbf r}_{\rm A \leftarrow B}     }{c}\right] ^{-1}=
\left[ 1+\frac{   \dot{r}_{ {\rm rel,A} }     }{c}\right] ^{-1}
\notag
\end{equation}
\begin{equation}
    =  \left[ 1+ \frac {2 \pi}{ P_{\rm orb}}  \ \frac  { a_{\rm rel} \ e} {\ c \sqrt{1-e^2} }
  \sin \theta(t) \right] ^{-1},
\end{equation}
 where $ \vec{  \mathbf v}_{ {\rm rel,A} } $ and $ \dot{r}_{ {\rm rel,A} }  $ are, for the
 relative orbit  of PSR A about PSR B,  the vector velocity and its radial (with respect to B) part, respectively.

\subsubsection{Apparent frequencies at the solar system barycentre}

In this section, we discuss the  transformations necessary to find apparent
frequencies of A's two beams at a given time  at the solar
system barycentre\footnote{It is standard procedure for pulsar observers to remove
the effects of the Earth's position and motion from pulsar periods,
frequencies,  and arrival times by reducing them
to their equivalent values at the solar system barycentre.},
in terms of  emission
times\footnote{ A desired emission time can be determined as a function of  ssb time,
by calculating the propagation time between the two locations.   See
\S\ref{sec:algorithm}
for details. },
Doppler factors, intrinsic orbital and spin frequencies.  Table 2 displays detailed
expressions associated with these items.

{\it{The first (direct) beam from A:}} The  intrinsic frequency of
this beam,  $f_{\rm A,0}$, will
be  Doppler-shifted upon receipt at the solar system
barycentre due to  A's orbital velocity at the time of emission, yielding
$f_{\rm 1,ssb}=f_{\rm A,ssb} =  {\mathbf{D}}  [ {\rm A,ssb} ] \times f_{\rm A,0}  $.

{\it{The second (``two-legged'') pulsed signal from A:}} This signal travels  along
two  segments in sequence:  first from A to B and then
from B to the solar system barycentre.  Some investigators
(e.g., \citet{Freire09a} )
suggest that this
``signal'' may consist of relativistic charged particles rather than photons,
but either will behave similarly in our frequency-based analysis. (While
the arrival-time
model of \citet{Freire09a} allows for a possible longitude offset between A's
radio and electromagnetic beams, this constant {\it{phase}} offset drops out of
our frequency-based analysis.)

The first leg's pulsed signal will be received
and  reemitted by B as a
``modulation'' with an apparent frequency in B's frame of $f_{\rm m\ at\  B\ due\ to\ A}$.
This frequency is  modified from its intrinsic value  $f_{\rm A,0}$, primarily by A's orbital
motion about B in a
fashion analogous to the modification of Earth's intrinsic spin frequency
to its solar one,
as described in Section 2. This requires that

\begin{align}
f_{\rm m\ at\  B\ due\ to\ A} =
\; \;\ \; \;\ \; \;\ \; \; \; \;\ \; \;\ \; \;\ \; \; \; \;\ \; \;\ \; \;\ \; \; \; \;\ \; \;\; \;\;\\
 {\mathbf{D}} [{\rm A,B}] \times
                          \begin{cases}
               ( f_{\rm A,0}  - f_{\rm orb}(t)\  )&\text{prograde spin,} \\
               ( f_{\rm A,0} + f_{\rm orb}(t)\  )&\text{retrograde spin,}  \\
                \ f_{\rm A,0}                           & \text{pulsation (no}     \\
                 & \text{\ \ \ \ \ \ \ \ \ \ \  \ spin);} \notag\\
           \end{cases}
 \label{eqn:intrinsicmod}
\end{align}
as summarized in Table 2.

 It is easy to show that this kinematic consequence of
the lighthouse model leads to exactly one fewer (prograde case) or one extra (retrograde case)
pulse from PSR A per orbital period, compared with the no-spin number. The former case is closely
analogous to the (prograde) Earth - Sun system, where  there is one fewer
solar day than sidereal days per Earth year.

Eq.  6
 would have a particularly simple form
 if the orbit were circular, both
because $f_{\rm orb}$ could be treated as a constant and because the
Doppler shift function  $ {\mathbf{D} } [{\rm A,B}] \equiv 1$ in this case.
The difference between the exact, elliptical orbit expression given in Eq. 6
and this fictitious, circular-orbit expression   is intimately related to the difference
between the apparent and
mean spin frequencies of the Earth with respect to the Sun
which
leads to a time-varying difference between apparent and mean solar time,
dubbed the ``Equation of Time'' \citep{s92}.

The pulse train on the second portion of this trajectory, leaving B with a frequency
$f_{\rm m\ at\  B\ due\ to\ A}$ of Eq. 6,
will be Doppler shifted upon its reception at the
solar system barycentre  due B's orbital motion, leading to modulation frequency
\begin{align}
f_{\rm 2, ssb}=f_{\rm m, ssb} = {\mathbf{D}} [{\rm B, ssb}] \times  f_{\rm m\ at\  B\ due\ to\ A}
\;\;\;\;\;\;\;\;\;\;\;\;\;\;\;\;\;\;\;\;\;\;\;\; \tag{6a} \\
                         = {\mathbf{D}} [{\rm B, ssb}] \times  { \mathbf{D}} [{\rm A, B}] \times
                          \begin{cases}
               ( f_{\rm A,0}  - f_{\rm orb}(t) \ ) &\text{prograde spin,} \\
              (  f_{\rm A,0} + f_{\rm orb}(t) \ )&\text{retrograde spin,}  \\
                \ f_{\rm A,0}                           & \text{pulsation (no}\\
                 & \text{\ \ \ \ \ \ \ \ \ \ \  \ spin).} \tag{6b}\\
           \end{cases}
 \label{eqn:f2ssb}
\end{align}

\subsection{Spin-A-Induced Offsets in B's modulated signal}

\begin{figure}
\includegraphics[width=80mm ] {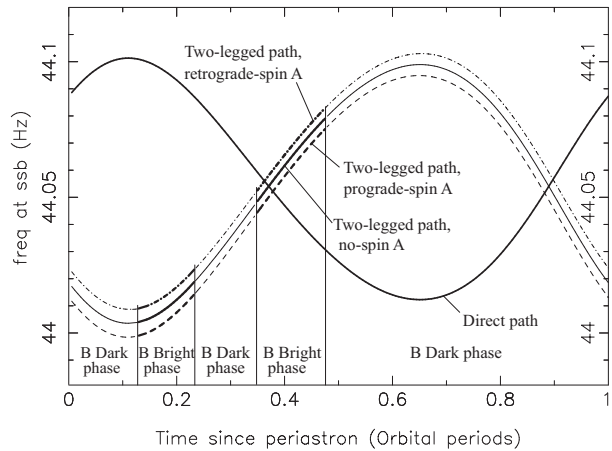}
\caption{Apparent frequencies of  direct and two-legged pulses originating
from  A, at the solar
system barycentre,   as a function of time for MJD 54155.  The  frequency of the direct path (A $\rightarrow$ ssb ),
$f_{\rm 1,ssb}$, is the isolated curve, while the three possible values of the
modulation pulse train frequency,
$f_{\rm m,ssb}=f_{\rm 2,ssb}$,
of the ``two-legged'' path (A $\rightarrow$ B $\rightarrow$ ssb)
 are depicted (with a $40 \times$ exaggeration of their
offsets from the middle, no-spin curve) as three closely-spaced curves.
Vertical lines delimit the parts of the orbit where B is visible (labelled ``Bright phase'') or invisible
(labelled ``Dark phase'').
Since the two-legged path is actually detected as a modulation of B's signal, the two-legged
path could never be observed during B's ``dark phases,'' so the two-legged curves are shown as
fainter lines at these phases.
B's modulation was actually only detected in a small region of the first bright phase,
although we believe it would be worthwhile to use our techniques to
search for modulations {\it{throughout both}} bright phases.   A similar,but mirror-
image, set of curves centered near  0.36 Hz would be created if a similar
pair of beams originates
 at PSR B (see \S\ref{sec:rotB}).
}
\label{fig:curve}
\end{figure}

The essence of our test for the rotation of A is contained in Eqs. 6,
since the sense or even the existence of A's spin  will select
one of the three versions of Eqs. 6b for the value of B's drifting-
subpulse-like modulation
frequency.  Hence the modulation has only three allowed values at
a given orbital phase. Fig. \ref{fig:curve}
  illustrates the  direct-path and the three possible two-legged-path
  pulse train frequencies   of the  beams from A
 at the solar system barycentre  as a function
of  time, with the offset of the two extreme  values
from the central, no-spin, curve for the two-legged beam
exaggerated by a factor  of forty.

Indeed, B's pulses are clearly modulated by A's emission
\citep{McLaughlin04a}.  Therefore, as long as the data are sufficiently
ample, the modulation frequency $f_{\rm m,ssb}$ should be obtainable with enough
precision to distinguish among the three possibilities.

As can be ascertained from Eqs. 6b,
 the mean  frequency offset\footnote{The instantaneous value,  also derivable from
 Eqs. 6b, varies about the mean owing to the ellipticity of the orbit, as discussed above.}
at the ssb between each of the three possible two-legged paths, is
\begin{equation}
| \Delta f_{\rm 2,ssb}| = | \Delta f_{\rm m,ssb}| = {\bf{D}}[{\rm B,ssb}] \times f_{\rm orb},
\end{equation}
while the mean {\it{fractional}} offset  $\delta$ is$^3$:
\begin{align}
&\delta
=  \frac{   | \Delta f_{\rm 2,ssb}|} {f_{\rm 2,ssb,no-spin}}
= \frac{   | \Delta f_{\rm m,ssb}|} {f_{\rm m,ssb,no-spin}}
= \frac{  {\bf{D}}[{\rm B,ssb}] \times f_{\rm orb} } {  {\mathbf{D}} [{\rm B, ssb}] \times  f_{\rm A,0} }  \notag \\
&= \frac{   f_{\rm orb}  }   {   f_{\rm A,0}  }.
\end{align}

The value of $ \delta$ is
only $0.00026\%$ [using  $f_{\rm orb} = 1 / (8834.535$  s) and $f_{\rm 0, A}=44.0540694$
Hz  \citep{Kramer06}].  While the three possible values of $f_{\rm m,ssb}$ are thus quite similar,
we note that they are precisely calculable via Eqs. (6b), as illustrated
in Fig.  \ref{fig:curve}. The next section details an algorithm for determining
the presence and sense of rotation of A
from a real data set, by finding which of the three modulation frequencies best matches the data.

\subsection{The algorithm }
\label{sec:algorithm}

The results of the above sections derive and illustrate the three possible
values of the ssb
modulation frequency at any given orbital phase. However, these frequencies vary
{\it{around}} the orbit due to Doppler shifts
and orbital ellipticity, thereby rendering  frequency searches difficult
without special techniques.  We discuss such techniques below.

Consider the observed, sampled (though incompletely) $k^{\rm th}$ intensity $I [k]$, measured at
solar system barycentric time $t[k]$:
\begin{equation}
t [k]= t_0 + \frac{k} { f_{\rm s} },
\end{equation}
where $ t_0$ is the ssb time of sample zero
and $f_{\rm s}$ is the sampling frequency.

 Full power spectral analyses could be  employed on chunks of this time
 series whose durations are short
enough to avoid significant spectral smearing  due to Doppler shifts
and orbital ellipticity.  The
highest-amplitude channel of
the power spectrum near the expected modulation frequency could then be
defined as the best estimate
of the modulation frequency for a given chunk;  and its
value could then be compared
 with each of the three
values of $f_{\rm m}$ predicted for that chunk  by Eqs. \ref{eqn:f2ssb},  thereby determining the
presence and sense of rotation of A. However, since the three expected values are so
 similar, it would  be difficult to distinguish among them in a relatively short and
 noisy chunk.

Therefore, it would be helpful to  eliminate the Doppler
and elliptical-orbit smearing of the modulation frequency around the orbit and
beyond, in order to employ much longer data sets.    Here we illustrate an algorithm to
resample the full extant data set
 in order to ``freeze''
the desired
modulation at a constant frequency, whose value can be determined via Fourier analysis of all the
data at once.  This procedure will result in much greater spectral resolution and
signal-to-noise.  The process is similar to procedures used in frequency-based searches of time
 series data for unknown binary
 pulsars, wherein the Doppler shifts due to putative orbital motion are eliminated
 via resampling of the data (e.g., \citet{allen13}).
However, while unknown binary searches require resampling the data with
multiple trial sets of parameters spanning an enormous search space,
our procedure requires only one parameter set for each of the three possible rotational states,
 all of which are well-determined.  In order
to resample the data, we first calculate times $t_{\rm B}[k]$
and $t_{\rm A}[k]$, the times at
B and A corresponding to ssb time $t[k]$ for the two-legged path from A to B
to the ssb that is responsible for the drifting-subpulse-like modulation.  Then
we derive an expression for the {\it{phase}} of the modulation  at B, taking
elliptical orbital motion into account.  Finally, we  show how to determine the presence and nature of
A's spin by searching for periodicities in this modulation phase space.

\subsubsection{ Freezing the frequency of B's pulses }

First, we calculate $t_{\rm  B}[k]$, the time of the $k^{th}$ sample measured at B,
by correcting for B's  propagation contributions along  the second leg, the path  from B
  to the ssb:
\begin{equation}
t_B[k] =t[k]  -\frac{L}{c} -\frac{z_{\rm B}} {c} = t_0+\frac{k}{f_{\rm s}} -\frac{L}{c}  -\frac{z_{\rm B}} {c};
\label{eqn:resampB}
\end{equation}
with $L$  the distance between the ssb and
binary barycentre bb\footnote{The unknown distance $L$ can be absorbed into $t_0$ by
an appropriate redefinition, with no loss of generality.};
and  $ z_{\rm B}$, the projection of
the
position of B with respect to the bb, $\vec{\mathbf r}_{\rm B}$,  onto the line of
sight $\hat{\mathbf r}_{\rm bb\leftarrow ssb} $:
 \begin{equation}
z_{\rm B} \equiv  \hat{\mathbf r}_{\rm bb\leftarrow ssb}  \cdot  \vec{\mathbf r}_{\rm B}=
\frac  {a_{\rm B} \sin  i \  (1-e^2) \sin( \omega_{\rm B}+\theta) }  {c \ (1+e \cos \theta)}.
\end{equation}
[See  Sec. \ref{sec:radialvel} for additional definitions, and \citet{Roy05} for a derivation.]

Eq. 10  can be used to transform $I[k]$, the  intensity data sampled at the ssb,  into a
resampled space  where the   previously time-variable
Doppler shift  $\mathbf{D} [\rm B, ssb]$  caused  by B's orbital motion is removed.  In this new
space, the pulses from B will yield a  { \it{fixed}}  power spectral peak   at $f_{B, 0}$.
Indeed, it is by this very technique
that  survey data are resampled in an effort to search for a binary pulsar of given orbital properties [e.g.,
 \citet{allen13}] (in this case, properties matching B's).

\subsubsection{ Freezing the frequency of the modulation }

In order to freeze frequency of the the {\it{modulation}} occuring when A's pulses
interact with B,
we must also   calculate $t_{\rm  A}[k]$, the time of the $k^{th}$ sample measured at A,
by correcting additionally for   propagation time  $|d_{\rm AB}| / c$  along  the
path  from A to  B:
\begin{equation}
t_A[k]=t_B[k]  -  \frac{|d_{\rm AB}|   }{c}
=   t_0+ \frac {k} { f_s}  -  \frac{L}{c} -\frac{z_{\rm B}} {c}   - \frac{|d_{\rm AB}|   }{c},
\label{eqn:tA}
\end{equation}
with [see also \citet{Freire09a}]:
\begin{equation}
\frac{|d_{\rm AB}| }{  c}  \approx \frac{ |r_{\rm AB}|}{c}  + \frac   {dr_{\rm AB} }{dt}  \Delta t_{\rm AB}
=\frac{ |r_{\rm AB}|}{c}    (1+ \frac   {dr_{\rm AB} }{dt} ),
\label{eqn:tAB}
\end{equation}
where $|r_{\rm AB}|$ is the {\it{instantaneous}} separation between the two pulsars.

Eq. \ref{eqn:tA} implicitly  eliminates the Doppler factors $\mathbf{D} [\rm B, ssb]$
and $\mathbf{D} [\rm A, B]$  in Eq.  6b caused by both pulsars'  radial
motions along the two-legged path.

Now let us more carefully specify the (``unprime'') sidereal reference frame centred on A as one
whose  x-axis points from A to B's position at periastron, and whose  y-axis points
in the direction of B's motion at periastron.    A is pulsing, or else
spinning in a prograde or retrograde manner with
  respect to   B's revolution, all at a rate whose magnitude is $f_{\rm A,0}$\footnote{For a
  sufficiently long time series,
  $f_{\rm A,0}$ must be
  a function of time to  account for pulsar A's spindown.}
  ({\it{cf.}} Table \ref{tab:one}),   measured in this unprime  frame.

Therefore, the rotational or pulsational phase of A in the unprime,
 sidereal frame centred on A would be essentially$^5$
  strictly proportional to $t_{\rm A}:$
   \begin{align}
\Phi_{\rm A,0}[k]=
           \begin{cases}
               + 2\pi f_{\rm A,0} \ t_{\rm A}[k] &\text{prograde spin.} \\
                - 2\pi f_{\rm A,0} \ t_{\rm A}[k] &\text{retrograde spin.}  \\
               \ \ 2\pi f_{\rm A,0} \ t_{\rm A}[k] & \text{pulsation (no spin).}
           \end{cases}
\label{eqn:Aphases}
\end{align}

While $f_{A,0}$ is the sidereal frequency of A's pulsations or rotations,
it is not necessarily  the frequency of A's signal's interception and modulation of B (i.e., it is not
necessarily equal to the modulation frequency).  Specifically,
if A emits a {\it{rotating}} lighthouse beam, a correction must be made due to B's orbital motion.
(Conversely, if A is instead {\it{pulsating}}, no such correction is
necessary because the pulsation is presumed to be emitted simultaneously in all directions.

In order to correct for orbital motion in the case of a rotating lighthouse beam,
consider  a second (``prime'') two-dimensional coordinate system lying in the orbital plane
 and centred on A, whose axes \{x',y'\},
 rotate (nonuniformly, due to the elliptical orbit)  such that its x'-axis points from A toward B's
 location at time   $t_{\rm B}[k]$.   The phase angle between the two coordinate systems (e.g.,
 between the x- and x'-axes) is
 just the true anomaly  $\theta(t_{\rm B}[k]$ ({\it{cf.}}  Sec. \ref{sec:radialvel}).

 The drifting-subpulse-like  modulations, which are created whenever A's
emission intercepts B, occur at intervals separated by exactly one rotation or pulsation of A
{\it{in the prime frame.}}    Therefore, we can write an expression for the corresponding
``modulation phase'' $\Phi_{\rm m \ at\ B\ due\ to\ A}[k]$,
which is just the rotational or pulsational phase of A in the {\it{prime}} frame:

\begin{align}
\Phi_{\rm m \ at\ B\ due\ to\ A}[k]=
\;\;\;\;\;\;\;\;\;\;\;\;\;\;\;\;\;\;\;\;\;\;\;\;\;\;\;\;\;\;\;\;\;\;\;\;\;\;\;\;\;\;\;\;  \notag\\
           \begin{cases}
               + 2\pi f_{\rm A,0} \ t_{\rm A}[k] - \theta(t_B[k]) &\text{prograde spin,} \\
               - 2\pi f_{\rm A,0} \ t_{\rm A}[k] - \theta(t_B[k]) &\text{retrograde spin,}  \\
                +2\pi f_{\rm A,0} \ t_{\rm A}[k]                  & \text{pulsation (no spin).}
           \end{cases}
\label{eqn:modphases}
\end{align}

The  modulations now occur at phase intervals of
$\Delta\Phi_{\rm m \ at\ B\ due\ to\ A}[k]= 2 \pi j$, with $j$ any positive integer.
In this fashion, we have
achieved our ultimate goal of ``freezing'' the {\it{modulations}} at a fixed periodicity in this transformed space,
even in the presence of ellipticity-induced, time-variable orbital frequencies.

An observer can distinguish only the {\it{magnitude}} of the modulation phase,
 $ | \Phi_{\rm m \ at\ B\ due\ to\ A}[k]|$, with
 \begin{equation}
 | \Phi_{\rm m \ at\ B\ due\ to\ A} [k] |_s= 2\pi f_{\rm A,0} \ t_{\rm A}[k] -s\  \theta(t_B[k]) ,
 \label{eqn:magmodphase}
\end{equation}
where A's spin state $s=\{+1,-1,\ 0\}$ for \{prograde spin,
retrograde spin, no spin but pulsation\}, respectively.  (Note that
 the right side of Eq. 16 will always be positive, as desired, since
 $f_{\rm A,0} >>d \theta / dt$.)

\subsubsection{Determining A's spin state  s from observations   }

 We have   expressed  the modulation
  phase magnitude at B, of the $k^{th}$ sample for
 each of the three possible spin states in Eq \ref{eqn:magmodphase}. We show below that this enables us to
  measure the  modulation  {\it{frequency}} at B
  via  a periodicity search,
thereby determining A's true spin state.

In order to search for modulation periodicities over a {\it{range}} of frequencies,
 we will now represent  the magnitude of A's sidereal rotation period by a generalized variable
 $f_{\rm A, trial}$, whose  value  will be   allowed to vary slightly about its known value:
   \begin{equation}
  f_{\rm A, trial} = z  f_{\rm A,0},
   \end{equation}
where the frequency search factor $z \approx 1$.
Then for each chosen $z$ and $s=\{+1,-1,0\}$,
the modulation phase magnitude at B of the $k^{th}$ ssb sample is
 \begin{equation}
 | \Phi_{\rm m \ at\ B\ due\ to\ A}[k,z]|_s=
                2\pi \ ( z\ f_{\rm A,0})\ t_{\rm A}[k] - s\  \theta(t_B[k])
\label{eqn:onemagmodphase}
\end{equation}

We can now test for the presence of  a modulation periodicity due to one of
the three possible spin states,
by doing a periodicity search in the vicinity of each such state
in modulation-phase space.
We associate the relevant modulation phase factor for a given state $s$, with each
ssb-sampled intensity $I[k]$, and  prepare  a Fourier
power spectrum of the product. The resulting quantity $P_{\rm n}(z   f_{\rm A,0})_s$, the power in the
$n^{th}$ harmonic of the modulation corresponding to A's trial spin frequency
$z   f_{\rm A,0}$ and spin state $s$,  is given by
\begin{equation}
P_{\rm n}(z   f_{\rm A,0})_s=
\left |\sum_k I[k]\;\exp{(-i\ n\  |\Phi_{\rm m \ at\ B\ due\ to\ A}[k,z]  |_s) } \right |^2,
\end{equation}
where $k$ can extend over any subset of
the full ssb-sampled dataset.

The true spin state will then be manifested as a sharp peak in one of the
three power spectra at
the frequency corresponding to  $z \equiv 1$ and its harmonics,  while the power spectra generated for the
other two putative spin states will exhibit no such peaks at the expected frequency and its harmonics.
If necessary, the sought-after signal can be further enhanced by the process of harmonic
summing, which is a well-established pulsar search technique.

The theoretical framework applied here can be further verified with a few additional observational
procedures.  For example, since the  {\it{modulation}}  of B's pulses can occur only at those
orbital phases where B's pulses are present (denoted ``bright phases'' in Fig. 2), the sought-after modulation {\it{periodicity}} must also  be visible only at those phases as well.
[Indeed, the modulation phenomenon is only directly visible near the
{\it{beginning}} of the first bright phase \citep{McLaughlin04a}, although we believe it is
worth  deploying our  technique with multiple trials on data from a wide range of bright phases.]
Conversely, while the pulses arriving {\it{directly}}
from A are observable at all orbital phases, the above  procedures should
 filter out their presence in the
resampled data.This ability to filter  for the direct signal is  key,
especially if a pulsation (no spin) result appears. The success of
the filtering out of the direct ray from A can itself be assessed by using our technique only on the
(B-) dark phases, where only the direct A ray is emitted.

\section{ An alternative: Determining the presence and sense of rotation of B }
\label{sec:rotB}
An  analysis of the apparent pulse train frequencies on the two equivalent paths from
PSR B  to the ssb (rather than the earlier case of paths originating at A) leads to identical
expressions as above, except with subscripts ``A'' and ``B'' interchanged.

Indeed, \citet{McLaughlin04b} and \citet{Breton12} show that A's pulses are modulated by B.
However, the modulation has so far only been detected during the short
(partial!) eclipse of A by B.  Unfortunately, this modulation
signal cannot be used to determine  B's rotation sense, because
it arises from a different  mechanism and hence the
signal relationships differ from those derived above.

If A's  signal modulation
by B can also be found in the non-eclipse phase,
 B's rotation sense may be determined more easily than A's.  In this case, with
$f_{\rm 0, B}= 0.360560355 $ Hz  \citep{Kramer06},
the equivalent mean fractional frequency offset distinguishing the
three possible
spin states of B is a much larger $0.031 \%$.
The data segments corresponding to the eclipse phases should
be discarded in the analysis in order to avoid contamination from modulation
originating by a different mechanism.
While B's radio signal disappeared in 2008 due to relativistic precession of the
spin axis \citep{Perera10}, it is possible that its spinning
 magnetosphere could still modulate A's pulses.  In addition,
B's beam should eventually precess back into our line of sight. Moreover, it may be worth
using our techniques to investigate whether A's signal is modulated by B in the non-eclipse phase, although it is unlikely for B to modulate A's emission in the same way that A does B because the spindown power, which is proportional to
$f \times \dot{f}$, is much less for B than for A \citep{KramerWex2009}.

\section{ Conclusion}

\cite{McLaughlin04a} presented a modulation pattern similar to
drifting subpulses in the signal of PSR B, but with a frequency
of 44 Hz, close to the pulse frequency of PSR A.
 The presence and sense of rotation of
A is encoded in the observed modulation pattern, and can
be revealed through an arrival-time-based analysis \citep{Freire09a} or the
frequency-based analysis presented above.  Our procedure offers the benefits
of relative conceptual simplicity and close analogy with familiar
phenomena in the Earth-Sun system.
We present a frequency-based procedure,  building upon that used
in binary pulsar search software,
to distinguish among direct, retrograde, or no
 rotation of PSR A by searching synchronously
for one of the three possible modulation signals over the full span of  available data.

Although the lighthouse model has been widely accepted, there has
nevertheless been no direct observational evidence in its support up to this time.
A  strength of our technique is its ability to provide such a test, empirically supporting
or refuting the model.

 \citet{ferdman13} have shown that  A's spin and orbital axes are aligned to within
$3 \degr$, but they have no direct means of distinguishing parallel from antiparallel alignment.  By assuming
parallel alignment, they are able to conclude that the second supernova, which created B, was relatively
symmetric. Therefore the presence and sense of the rotation, as revealed by this analysis, will
further test their and others' (e.g., \citet{kramerstairs08,farret11}) evolutionary scenarios.

While \citet{McLaughlin04b} also found that A's signal is modulated at B's
frequency near A's eclipse, the mechanism is different and not dependent on
B's rotation.  If, however, the modulation is observed in the future away from
eclipse, it will provide an easier test for B's rotation than does the approach
delineated above for A.

Our analysis can also be used on other
binary systems discovered to possess phenomena caused by mutual interactions.

\section*{Acknowledgments}

JMW is supported by U.S. National Science Foundation Grants AST-0807556 and AST-1312843.
ZXL and YL are grateful to Xiang-Ping Li, Ali Esamdin, and Jian-Ping Yuan for very useful discussions.

\end{document}